\documentstyle[multicol,pre,aps,epsf]{revtex}
\begin{document}
\draft
\title{Dynamics of $\alpha$- and $\beta$-processes in thin polymer
films: \\ poly(vinyl acetate) and poly(methyl methacrylate)
}
\author{Koji Fukao\cite{A}, Shinobu Uno$^a$, Yoshihisa Miyamoto, 
Akitaka Hoshino$^a$, and Hideki Miyaji$^a$}
\address{
Faculty of Integrated Human Studies,
Kyoto University, Kyoto 606-8501,
Japan\\
$^a$Department of Physics,
Kyoto University, Kyoto 606-8502,
Japan }

\date{Received May 9, 2001}
\maketitle

\begin{abstract}

Dynamics of thin films of  poly(vinyl acetate) (PVAc) and 
poly(methyl methacrylate) (PMMA)
have  been investigated by dielectric relaxation spectroscopy in the
frequency range from 0.1Hz to 1MHz at temperatures from 263K to 423K.
The $\alpha$-process, the key process of glass transition, is observed
for thin films of PVAc and PMMA as a dielectric loss peak
at a temperature $T_{\alpha}$ 
in temperature domain with a fixed frequency.
For PMMA the $\beta$-process is also observed 
at a temperature $T_{\beta}$.
For PVAc, $T_{\alpha}$ 
decreases gradually with decreasing thickness, and the thickness
dependence of $T_{\alpha}$ is almost independent of the molecular
weight ($M_{\rm w}\le$2.4$\times$10$^5$). For PMMA, $T_{\alpha}$
remains almost constant as thickness decreases down to a critical 
thickness $d_c$, at which
point it begins to decrease with decreasing thickness. Contrastingly,
$T_{\beta}$ decreases gradually as thickness decreases  to $d_c$, 
and below $d_c$ it
decreases drastically. For both PVAc and PMMA, the broadening of the
distribution of the relaxation times in thinner films is observed and
this broadening is more pronounced for the $\alpha$-process than for
the $\beta$-process. It is also observed that the relaxation strength
is depressed as the thickness decreases for both the polymers.
\end{abstract}\

\begin{multicols}{2}
\section{\bf INTRODUCTION}
Recent progress in theoretical and experimental studies has 
clarified many properties of glass transitions~\cite{Vigo,YKIS}. 
However, the mechanism of the glass transition has not yet been 
fully understood~\cite{Angell}.
The major issue is the experimental approach to the 
investigation of the length scale of the glass transition and 
dynamics of the $\alpha$-process such as dynamical
heterogeneity~\cite{Ediger,Sillescu}. 
According to the Adam and Gibbs theory, the dynamics of glass 
transitions are associated with cooperative motions which are 
characterized by a so-called {\it cooperatively rearranging region} (CRR),
in which molecules move cooperatively with each other~\cite{Adam-Gibbs}.
The size of the CRR is assumed to increase as temperature decreases
to the glass transition temperature $T_{\rm g}$. The characteristic 
length scale associated with such cooperative motions 
has been investigated by multi-dimensional
NMR~\cite{Spiess}, dielectric hole burning~\cite{Schiener} and
photobleaching~\cite{Ediger2} for bulk systems. 
Another approach to studying the
length scale is the investigation of the finite size effect on glass
transition dynamics through confined systems such as thin polymer
films~\cite{Forrest1} or small molecules in nanopores~\cite{Confinement}.

The first direct measurement of the reduction in $T_{\rm g}$ in thin 
polymer films has been made by Keddie {\it et~al.} in 1994~\cite{Keddie1}. 
In their work,
the film thickness of the thin films of polystyrene (PS) supported on a 
hydrogen
passivated silicon wafer has been measured as a function of temperature
by an ellipsometer. The results showed that the observed $T_{\rm g}$ in
such thin films can be described as a function of film thickness, $d$,
as follows:
\begin{eqnarray}\label{Tg_Keddie}
T_{\rm g}(d) &=&T_{\rm g}^{\infty}\left[1-\left(\frac{A}{d}\right)^{\delta}\right],
\end{eqnarray}
where the best-fit parameters are $T_{\rm g}^{\infty}$=373.8K,
$A$=1.3$\pm0.1$nm, and $\delta$=1.28$\pm$0.20~\cite{Keddie3}. 
The molecular weight ($M_{\rm w}$) dependence of $T_{\rm
g}(d)$ has not been observed for $M_{\rm w}$ range from 1.2$\times 10^5$
to 2.9$\times 10^6$, and hence it was concluded that the confinement 
effect of polymer chains within thin layers may be neglected in such
systems. Further measurements of $T_{\rm g}$ in thin films of
poly(methyl methacrylate) (PMMA) supported on two different kinds of 
substrate revealed that the strong attractive interactions between 
substrate and polymers lead to increase in $T_{\rm g}$ with decreasing 
film thickness~\cite{Keddie2}. 

In order to remove such interactions between substrate and polymers,
Forrest {\it et al.} measured $T_{\rm g}$ for {\it freely standing films} of 
polystyrene by Brillouin light scattering 
measurements~\cite{Forrest2,Forrest3}. They obtained the thickness 
dependence of $T_{\rm g}$ as follows:
\begin{eqnarray}\label{Tg_Forrest}
T_{\rm g}(d) =\left\{ \begin{array}{c@{\quad:\quad}l} 
T_{\rm g}^{\infty}\left(1-\frac{d_0-d}{\zeta}\right) & d < d_0 \\
T_{\rm g}^{\infty} & d > d_0,
\end{array}\right.
\end{eqnarray}
where $T_{\rm g}^{\infty}$ is the glass transition temperature for thick
films, $\zeta$ is the constant and $d_0$ is the critical thickness.
The critical thickness $d_0$, below which $T_{\rm g}$ decreases linearly
with decreasing film thickness, depends on the molecular weight,
$i.e.$, $d_0$ increases with the molecular weight in a way similar 
to the radius of gyration of polymer chains.

In our previous papers, we performed electric capacitance measurements
for thin films of polystyrene supported on Al-deposited glass substrate
and determined the $T_{\rm g}$ as a temperature at which the temperature
dependence of capacitance changes discontinuously~\cite{Fukao1,Fukao2}. 
As a result, we confirmed the 
reduction of $T_{\rm g}$ in our system in a way similar to 
Eq.(\ref{Tg_Keddie}). Furthermore, we made volume relaxation
measurements
on thin polystyrene films and measured the temperature $T_{\alpha}$
at which the imaginary component of complex thermal expansion
coefficient has a peak at a given very low frequency corresponding to 
the relaxation time about 100 sec during the heating process of a constant
rate~\cite{Fukao3,Fukao4}. 
The temperature $T_{\alpha}$ thereby obtained can 
be regarded as the glass transition temperature $T_{\rm g}$. In our
measurements it is found that $T_{\rm g}$ decreases
slightly with decreasing film thickness down to a critical thickness
$d_c$, and below $d_c$ it decreases very rapidly. The value of $d_c$
changes with $M_{\rm w}$ in the similar way to the radius of gyration
of polymer chains. This behavior of $T_{\rm g}$ is quite similar to 
those obtained in freely standing films as shown in
Eq.(\ref{Tg_Forrest}) except the slight change in $T_{\rm g}$ above
$d_c$.

Intensive investigations have been performed so far mainly on 
polystyrene films~\cite{Forrest1}. It is important to 
clarify whether the results
extracted from the investigations on thin polystyrene films hold 
also for other polymers. In this paper, therefore, we investigate 
the dynamics
of thin films of poly(vinyl acetate)(PVAc) and PMMA by dielectric 
relaxation spectroscopy.
PVAc is a polymer  suitable for dielectric relaxation measurements
because a monomer unit of PVAc has a large dipole moment~\cite{McCrum}.
In PMMA, there is a strong $\beta$-process which is due to
the hindered rotation of the side group in addition to the
$\alpha$-process~\cite{McCrum}. 
It is indispensable to investigate the 
size dependence of dynamics of the $\alpha$-process and the
$\beta$-process when elucidating the nature of glass transition.

This paper consists of five sections. After the introduction,
experimental details on preparation method of thin films and dielectric
relaxation spectroscopy 
are given in Sec.II. In Sec.III, experimental
results on poly(vinyl acetate) and poly(methyl methacrylate) by
dielectric relaxation spectroscopy are shown. 
In particular, thickness dependence of the temperatures $T_{\alpha}$ and
$T_{\beta}$ corresponding to the $\alpha$- and $\beta$-peaks in the
dielectric loss and the width of the distribution of the relaxation time
of the $\alpha$- and $\beta$-processes are focused in Sec.III.
Discussions on the present 
experimental results compared with the results reported previously are
given in Sec.IV and a summary is given in Sec.V.

\vspace*{-0.2cm}
\section{\bf Experiments}
The polymer samples used in this study (PVAc and PMMA) are purchased
from Scientific Polymer Products, Inc. The molecular weight and radius
of gyration of the polymers are listed in Table I.
The glass transition 
temperatures of PVAc and PMMA in the bulk states are 303K and 373K,
respectively. Thin polymer films were prepared by spin-coating a toluene 
solution of PVAc (PMMA) onto Al-deposited glass substrate. 
Film thickness is controlled by changing the
concentration of the solution. The thin films obtained by spin-coat
method were annealed {\it in vacuo} for 48 hours at 303K for PVAc and 
353K for PMMA,
respectively. After annealing, Al was vacuum-deposited once more onto the
thin films to serve an upper electrode.

Dielectric measurements were performed by using an LCR meter (HP4284A)
for the frequency range from 20Hz to 1MHz and an impedance analyzer
(Solartron Instruments SI1260) for the frequency range from 0.1Hz to
1MHz. The temperature of a sample cell was changed between 273K and
373K for PVAc and between 263K and 423K for PMMA at a constant rate of
0.5K/min. The dielectric measurements during the heating and cooling
processes were performed repeatedly several times.
Data acquisition was made during the above cycles except the first
cycle. The good reproducibility of dielectric data was obtained 
after the first cycle.

The thickness $d$ is related to the electric capacitance $C'$ of thin
films in the
following way: $C'=\epsilon'\epsilon_0\frac{S}{d}$, where $\epsilon_0$
is the permittivity of the vacuum, $\epsilon'$ is the permittivity of 
the polymer (PVAc or PMMA) and $S$ is the effective area of the
electrode ($S$=8.0mm$^2$).  For the frequency range where there are no
contributions due to any dielectric dispersion, $\epsilon'$ can be
regarded as constant for any change in frequency, and hence the film 
thickness is
inversely proportional to the electric capacitance. Relative film
thickness at a given temperature can be obtained from the electric 
capacitance of the thin films
at a frequency within the above frequency
range. The frequency and the temperature we chose for the determination
of the relative thickness are 8kHz and 298K for PVAc, and 1kHz and 273K for 
PMMA. The absolute values of the film thickness for several films were
measured directly  by an atomic force microscope (Shimadzu SPM-9500) 
in order to calibrate the film thickness.

As shown in a previous paper~\cite{Fukao2}, the resistance of the Al 
electrodes cannot 
be neglected for dielectric measurements of very thin films. This
resistance leads to an artifact loss peak on the high frequency side;
this peak results from the fact that the system is
equivalent to a series circuit of a capacitor and
resistor~\cite{Kremer}. Because the
peak shape in the frequency domain is described by a Debye-type
equation, the ``C-R peak'' can easily be subtracted. The data thus
corrected were used for further analysis in the frequency domain.

\section{\bf Results}
\subsection{\bf Poly(vinyl acetate)}
\subsubsection{\bf Dielectric relaxation of the $\alpha$-process in PVAc}
Figure 1 shows the dependence of the complex electric capacitance 
($C^*=C'-iC''$) on the logarithm of frequency at various temperatures
for thin films of PVAc with $M_{\rm w}$=1.8$\times$10$^5$ and film
thicknesses (a) $d$=440nm and (b) $d$=16nm. In this figure, we find that 
the imaginary component $C''$ has a peak due to the $\alpha$-process. 
The peak frequency shifts to the higher frequency side as
temperature increases. Comparing Fig.1(a) and Fig.1(b), we find that the 
peak position shifts to the higher frequency side at a fixed
temperature as the film thickness decreases from 440nm to 16nm.

The temperature change in the dielectric loss at 100Hz
normalized with the peak value for the bulk sample is shown in Fig.2 in 
the case of PVAc with
$M_{\rm w}$=1.8$\times$10$^5$. The dielectric peak due to the $\alpha$-process 
possesses a
maximum at the temperature $T_{\alpha}$. It is found in Fig.2 that the 
temperature $T_{\alpha}$ decreases with decreasing film thickness. 
At the same time the peak width of the $\alpha$-process increases and the
height at $T_{\alpha}$ decreases with decreasing film thickness. This
behavior is related to the change in the distribution of the relaxation 
times and the relaxation strength with the film thickness. 

Figure 3 displays the temperature dependence of the frequency $f$ of the 
$\alpha$-process for thin films of PVAc with three different 
thicknesses 440nm, 62nm, and 18nm, where $f$ is associated with 
a characteristic time of the $\alpha$-process $\tau_{\alpha}$ via the relation
$2\pi f\tau_{\alpha}$=1. 
Each point in Fig.3 consists of a data set ($1/T_{\alpha}$, $\log_{10} f$).
For each film thickness, the temperature dependence of $\tau_{\alpha}$
can well be reproduced by the Vogel-Fulcher-Tammann (VFT)
law~\cite{VFT1}: $\tau_{\alpha}(T)=\tau_{\alpha}^{\infty}\exp [U/(T-T_0)]$, 
where $\tau_{\alpha}^{\infty}$ is the relaxation times at very high
temperatures, $U$ is the apparent activation energy, and $T_0$ is the
Vogel temperature. 
At a given temperature, the frequency $f$ 
increases with decreasing
film thickness, $i.e.$, the relaxation becomes faster in thinner
films. This thickness dependence of $f$ (and $\tau_{\alpha}$) becomes 
stronger as
the temperature approaches the glass transition temperature. 
This behavior is quite similar to that observed in thin films of PS
supported on glass substrate~\cite{Fukao1,Fukao2}.

In order to check how the $\alpha$-process changes with film thickness, 
the thickness dependence of the temperature $T_{\alpha}$ at the
frequency 100Hz is shown in Fig.4 for three different molecular weights 
 1.2$\times$10$^5$, 1.8$\times$10$^5$ and 2.4$\times$10$^5$.
The error bars in Fig.4 stand for  the standard deviation in
$T_{\alpha}$ for the data acquired by measurements done repeatedly several
times. Because the glass transition temperature $T_{\rm g}$ can be
regarded as the value of $T_{\alpha}$ obtained for a very low frequency
of the applied electric field, for example, $f\approx 1/(2\pi\cdot 10^2$ sec),
the absolute value of $T_{\rm g}$ is lower than that of the $T_{\alpha}$
for $f$=100Hz.
However, 
it is possible to extract qualitative information on the thickness
dependence of $T_{\rm g}$ from the observed thickness dependence of 
$T_{\alpha}$ in Fig.4. 

The solid curve in Fig.4 is  obtained by using the same form of 
Eq.(\ref{Tg_Keddie}) in which $T_{\rm g}$ is replaced by $T_{\alpha}$
with parameters as follows: $T_{\alpha}^{\infty}$=334.3$\pm$0.1K,
$A=0.11\pm 0.03$nm, $\delta =0.77\pm 0.04$. Because the single curve can
reproduce the observed values of $T_{\alpha}$ for the three different
molecular weights, the temperature $T_{\alpha}$ seems to have almost 
no $M_{\rm w}$ dependence 
and hence we conjecture that $T_{\rm g}$ is also independent of the
molecular weight for $M_{\rm w}$=1.2$\times$10$^5$-2.4$\times$10$^5$ and 
$d>$10nm. 

In order to discuss the dynamics of the $\alpha$-process in thin
films of PVAc, the observed complex dielectric constant 
$\epsilon^*(\omega)$ in Fig.1 is fitted to the following model function:
\begin{eqnarray}\label{eps1}
\epsilon^*(\omega )=\epsilon_{\infty}+A\omega^{-m}e^{-i\frac{\pi}{2}m}
+\frac{\Delta\epsilon}{(1+(i\omega\tau_{0})^{\alpha_{\mbox{\tiny HN}}})^{\beta_{\mbox{\tiny HN}}}},
\end{eqnarray}
where $\omega =2\pi f$ and $\epsilon_{\infty}$ is the permittivity at
a very high frequency. The second term in the r.h.s. is a contribution 
from space charge~\cite{Miyamoto}, and it can be attributed to pure DC 
conductivity if $m=1$. 
In the case of PVAc, the fitted value of $m$ was found to be between 
0.40 and 0.93 
depending on film thickness. The third
term in the r.h.s. comes from the $\alpha$-process
and its form is empirically proposed by Havriliak-Negami~\cite{HN}, where
$\Delta\epsilon$ is the relaxation strength, $\tau_0$ is the
characteristic time, $\alpha_{\mbox{\tiny HN}}$ and $\beta_{\mbox{\tiny HN}}$ are the shape parameters.

The thickness dependence of the relaxation strength
$\Delta\epsilon$ obtained from the observed values of
$\epsilon^*(\omega)$ at 333K is shown in Fig.5 for thin 
films of PVAc with $M_{\rm w}$=2.4$\times$10$^5$, 
where $\Delta\epsilon$ is normalized with respect 
to the value of $\Delta\epsilon$ for the bulk sample (See open circles).
The temperature 333K was chosen because the dielectric loss peak of the
$\alpha$-process is 
located at the center of the frequency window in the present measurement 
at this temperature. 
As the film thickness decreases, the normalized relaxation strength decreases 
from 1 to about
0.2. This behavior in relaxation strength corresponds to the decrease in 
the height at $T_{\alpha}$ in dielectric loss in Fig.2. The $d$ dependence of 
$\Delta\epsilon$ will be discussed later.
The fitting parameters $\alpha_{\mbox{\tiny HN}}$ and 
$\beta_{\mbox{\tiny HN}}$ of the HN equation are shown 
in Fig.6. It is found that with decreasing film thickness the parameter 
$\alpha_{\mbox{\tiny HN}}$ decreases and $\beta_{\mbox{\tiny HN}}$ 
slightly increases.
The decrease in $\alpha_{\mbox{\tiny HN}}$ with decreasing film thickness implies the
broadening of the distribution of the relaxation times.

\subsubsection{\bf Distribution of the relaxation times}

The experimental observation that the $\epsilon^*(\omega)$ does not obey 
a simple Debye equation but rather a complicated HN equation can be
accounted for by assuming that $\epsilon^*(\omega )$ is expressed as the 
sum of the Debye equations with different relaxation times in the following
way:
\begin{eqnarray}\label{dist1}
\epsilon^*(\omega)=\epsilon_{\infty}+\Delta\epsilon\int^{+\infty}_{-\infty}
\frac{F(\log_e\tau) d(\log_e\tau )}{1+i\omega\tau },
\end{eqnarray}
where the second term in Eq.(\ref{eps1}) is omitted. $F(s)$ is a
distribution function of the logarithm of the relaxation times of the
$\alpha$-process ($s=\log_e\tau$) and is normalized as follows:
$\int^{+\infty}_{-\infty}F(s)ds=1$. If it is assumed that
$\epsilon^*(\omega)$ is expressed by the HN equation, 
the analytical form of $F(s)$ can easily be obtained in the following 
way with the parameters $\alpha_{\mbox{\tiny HN}}$, 
$\beta_{\mbox{\tiny HN}}$, and $\tau_0$:
\begin{eqnarray}\label{dist2}
F(s)&=&\frac{1}{\pi}(1+2e^{\alpha_{\mbox{\tiny HN}}(x_0-s)}
\sin\pi\alpha_{\mbox{\tiny HN}} +e^{2\alpha_{\mbox{\tiny HN}} 
(x_0-s)})^{-\beta_{\mbox{\tiny HN}}/2}\nonumber \\
&\times&\sin\left[\beta_{\mbox{\tiny HN}}\tan^{-1}\left(\frac{
e^{\alpha_{\mbox{\tiny HN}}(x_0-s)}\sin\pi\alpha_{\mbox{\tiny HN}} }
{1+e^{\alpha_{\mbox{\tiny HN}}(x_0-s)} \cos\pi\alpha_{\mbox{\tiny HN}} 
}\right)\right],
\end{eqnarray}
where $x_0=\log_e\tau_0$. 

Figure 7 displays the distribution of
$\log_{10}\tau_{\alpha}$ calculated in terms of Eq.(\ref{dist2}) with
the parameters $\alpha_{\mbox{\tiny HN}}$, $\beta_{\mbox{\tiny HN}}$ 
and $\tau_0$ 
at 333K 
for thin films of PVAc with $M_{\rm w}$=2.4$\times$10$^5$ and $d$=1400nm 
and 19nm. In this figure, it is found that the average relaxation time
of the $\alpha$-process decreases with decreasing film thickness, and
that the width of the distribution of the relaxation times
$\tau_{\alpha}$ becomes broader in thinner films. In order to
quantify the change in the distribution of $\tau_{\alpha}$, the full
width $w$ at the half maximum (FWHM) at 333K of $F(s)$ was evaluated
as shown in Fig.8(a) for thin films of PVAc with $M_{\rm
w}$=2.4$\times$10$^5$ (See open circles). It is found that the 
width $w$ increases with
decreasing film thickness. The thickness dependence of $w$ is given by
the equation $w(d)=w_0(1+a/d)$, where $w_0$=0.92$\pm$0.03 and 
$a$=22.0$\pm$1.6nm for the $\alpha$-process of PVAc. 
This broadening of the distribution of the relaxation times and its 
functional form have also been observed in thin films of polystyrene by
dielectric relaxation spectroscopy~\cite{Fukao2} and ellipsometric 
measurements~\cite{Kawana1}.

\subsection{\bf Poly(methyl methacrylate)}

\subsubsection{\bf Temperature dispersion of dielectric loss}
Figure 9 shows the temperature change in the imaginary part of the complex
dielectric constant at 
a fixed frequency for PMMA with $M_{\rm w}$=4.9$\times$10$^5$ 
and $d$=900nm and 9.5nm. 
The value of $\tilde\epsilon''$ is normalized with the maximum 
value for the bulk sample ($d$=900nm).
The value $\tilde\epsilon''$ at a
fixed frequency (40Hz) displays an anomalous increase with temperature due to
the $\alpha$- and $\beta$-processes, and it possesses two
maxima at the temperatures $T_{\alpha}$ and $T_{\beta}$.
The values of $T_{\alpha}$ and $T_{\beta}$ depend not only on the
frequency $f$ but also on the film thickness, as we can see in Fig.9; 
in the case of $f$=40Hz, $T_{\alpha}$=401K and $T_{\beta}$=320K for
$d$=900nm, and $T_{\alpha}$=383K and $T_{\beta}$=301K for $d$=9.5nm.
Both the height and the width of the $\beta$-peak are larger than those
of the $\alpha$-peak. The origin of the $\alpha$-process
is attributed to the micro-Brownian motion of the polymer main chain,
while that of the $\beta$-process is to the hindered rotation of
the side branch of PMMA~\cite{McCrum}.

In order to extract the peak temperatures $T_{\alpha}$ and $T_{\beta}$
and the widths $\Delta T_{\alpha}$ and $\Delta T_{\beta}$, the observed 
values $\epsilon''$ as a function of temperature were fitted by the sum of the
two Lorentzian functions 
$\tilde\epsilon_i''/\{1+[(T-T_i)/\Delta T_i]^2\}$
($i=\alpha, \beta$). The functional form has no physical meaning, but it 
is used only to obtain the peak temperature and the width. If necessary,
the contribution due to DC conductivity was also taken into account by
adding another peak at higher temperature to the fitting function. The
thickness dependence of the widths of the $\alpha$- and
$\beta$-processes in the temperature domain, normalized with those in
the bulk sample, $\Delta T_i/\Delta T_i^{0}$, is shown in Fig.8(b), 
where $\Delta T_i^0$ is the width for the bulk sample. 
The solid lines in Fig.8(b) are given by $\Delta T_i(d)$=$\Delta T_i^0
(1+a/d)$. In PMMA, $\Delta T_{\alpha}^0$=7.9$\pm$0.5K and
$a$=16.7$\pm$2.0nm for the $\alpha$-process and $\Delta T_{\beta}^0$
=48.4$\pm$0.9K and $a$=1.3$\pm$0.3nm for the $\beta$-process.

The dependence of the frequency $f$ on the reciprocal of the
temperatures $T_{\alpha}$ and $T_{\beta}$ is shown in the dispersion
map of Fig.10 for PMMA with $M_{\rm w}$=4.9$\times$10$^5$
with three different film thicknesses. The number of data points of the
$\alpha$-process is limited because the small $\alpha$-peak merges with
the large $\beta$-peak as the frequency increases, and as a result it is
much more difficult to extract $T_{\alpha}$ at higher frequencies.

As shown in Fig.10, the branch of the $\beta$-process is well separated
from that of the $\alpha$-process for the frequency range below about
500Hz. However, the $\beta$-process approaches the $\alpha$-process as
the frequency or the temperature increases, because the slope of the
straight line corresponding to the $\beta$-process is much smaller than
that of the $\alpha$-process: the corresponding apparent activation
energies in this frequency range  are 651$\pm6$ kJ/mol for the
$\alpha$-process and 32.7$\pm0.2$ kJ/mol for the $\beta$-process.
In this limited frequency range investigated here, thickness dependence 
of the apparent activation energy can not be observed.
(Here, it should be noted that the
$\alpha$-branch can indeed be fitted by a straight line in Fig.10,
because of the limitation of the frequency range, but we do not intend
to deny the validity of the VFT law.) In Fig.10 we find that both the 
$\alpha$- and $\beta$-processes shift to the higher frequency
side at a fixed temperature, as the thickness decreases, $i.e.$, both
the relaxations become faster. 

\subsubsection{\bf $T_{\alpha}$ and $T_{\beta}$ as a function of thickness}
In order to
quantify the thickness dependence of $T_{\alpha}$ and $T_{\beta}$, the
frequency is fixed to be 40Hz and the temperatures $T_{\alpha}$ and
$T_{\beta}$ are evaluated from the observed curve of $\epsilon''$
vs. $T$ for various film thicknesses from 9.5nm to 1000nm. The results
are summarized in Fig.11, where Figs. 11(a) and (b) display the thickness 
dependence of the temperatures $T_{\alpha}$ and $T_{\beta}$ observed at
40Hz for the thin films of PMMA with $M_{\rm w}$=4.9$\times$10$^5$,
respectively.

Figure 11(a) shows that $T_{\alpha}$ {\it remains almost constant} within the
experimental accuracy as the film thickness decreases down to a critical
thickness $d_{\alpha, c}$, at which point $T_{\alpha}$ begins to
decrease rapidly
with $d$. The solid curve in Fig.11(a) below $d_{\alpha,c}$ is given by a 
linear relation between $T_{\alpha}$ and $d$, although it does not
appear to be a straight line because of the logarithmic scale of the
horizontal axis. The amount of the depression of $T_{\alpha}$ in the
thin films is about 20K between the bulk sample and the thinnest film
with $d$=9.5nm. In this case, the critical thickness $d_{\alpha, c}$ is
found to be about 20nm for the $\alpha$-process at 40Hz.

On the other hand, thickness dependence of $T_{\beta}$ seems to be to
some extent different from that of $T_{\alpha}$, as shown in Fig.11(b).
The temperature $T_{\beta}$ {\it decreases slightly with decreasing film
thickness} down to a critical thickness $d_{\beta, c}$ even for the
relatively thick films. 
Below a critical thickness $d_{\beta, c}$, $T_{\beta}$ 
decreases more rapidly
with decreasing film thickness. The depression in $T_{\beta}$ is about 25K
also in this case for the thickness range investigated here, and the
critical thickness $d_{\beta, c}$ is about 19nm.
It is quite interesting that the critical thickness of the
$\beta$-process almost agrees with that of the $\alpha$-process;
$d_{\alpha,c}\approx d_{\beta,c}(\equiv d_c)$. This
result may suggest that there is a strong correlation between the
$\alpha$-process and the $\beta$-process, although the two processes at
40Hz exist quite separately on the dispersion map in Fig.10.

In order to investigate the dynamical behavior in thin films of PMMA
with $d$=900nm and 9.5nm, we 
observed the frequency dependence of the complex dielectric constant at
a fixed temperature, as shown in Fig.12. Corresponding to Fig.9, the
large peak due to the $\beta$-process can be seen in the 
$C''$ vs. $f$
plot and a small contribution from the $\alpha$-process can also be
observed at the foot of the large $\beta$-peak at 403K. According to
the procedure of data analysis done for PVAc, the observed $C^*$ is 
fitted to a model function consisting of the two HN functions. 
In Fig.12, the calculated values with this model function are shown as
solid curves. For thick films, the model function can very well 
reproduce the observed results for the temperature and frequency range
shown in Fig.12. For very thin films, the data at higher frequency
(above 10$^4$ Hz) include relatively large errors. Therefore, in such a
case,  we used the data below 10$^4$ Hz for the data fitting.

In order to evaluate the distribution $F(s)$ and the relaxation strength 
$\Delta\epsilon$, the fitting parameters were obtained for the 
$\beta$-process at the temperature
at which the $\beta$-peak is located at 300Hz for various film
thicknesses. In this case, because the temperatures are located between
323 K and 338 K depending on the film thickness, the contribution 
from the $\alpha$-process can safely
be neglected. Using the parameters $\alpha_{\mbox{\tiny HN},\beta}$, 
$\beta_{\mbox{\tiny HN},\beta}$,
and $\tau_{0, \beta}$ obtained by fitting to the HN equation, we evaluated the
distribution function $F(s)$ of the relaxation time of the
$\beta$-process for PMMA, where the subscript $\beta$ means that these 
parameters are obtained for the $\beta$-process,  and then 
we obtained the thickness dependence
of FWHM $w$. The result of $w$ is shown in Fig.8(a) 
(See open triangles).
For the $\beta$-process of PMMA, $w_0$=4.7$\pm$0.2 and $a$=3.20$\pm$0.02 nm.
In this figure, the result of $w$ for the $\alpha$-process in PVAc is
also shown instead of $w$ for the $\alpha$-process in PMMA, because 
it was difficult to obtain $w$ for the $\alpha$-process in PMMA. In
addition to $w$, the normalized widths of the dielectric loss peak 
in the temperature domain due to the $\alpha$- and $\beta$-processes, 
$\Delta T_i/\Delta T_i^0$ ($i$=$\alpha$, $\beta$), are
also shown in Fig.8(b). 
Because the relaxation time $\tau$ is a function of the temperature $T$,
the distribution function of relaxation times $F(\log_e\tau)$ can be
converted into that in the temperature domain, where the distribution
function of the relaxation times with respect to temperature is given by 
$\tilde F(T)\equiv F(\log_e\tau(T))\frac{d\log_e\tau}{dT}$.
Therefore, the distribution of the relaxation time can be evaluated not
only from $F(\log_e\tau)$ but also from the width in the temperature
domain.
Comparing these results with each other, we
found that the distribution of the relaxation times of the
$\alpha$-process becomes much broader with decreasing film thickness,
while for the $\beta$-process this broadening is weaker or remains
almost constant with decreasing film thickness.

Figure 5 shows the dependence of normalized relaxation strength on the
reciprocal of film thickness for the 
$\beta$-process in thin films of PMMA (See open triangles). 
The relaxation strengths are 
obtained by fitting the observed frequency dependence of dielectric loss
for the $\beta$-process at temperature at which the 
$\beta$-peaks are located at 300Hz. 
The dependence of normalized relaxation strength of the $\alpha$-process
in thin films of PVAc with $M_{\rm w}$=2.4$\times$10$^5$ is also 
shown in Fig.5. However, that of the $\alpha$-process in PMMA has not
yet been evaluated because the dielectric loss due to the
$\alpha$-process in this case is much smaller than that due to the
$\beta$-process. More precise and wider frequency range measurements
will be required to extract the relaxation strength of the
$\alpha$-process in PMMA especially in thinner films. 

In Fig.5,
it is found that the dielectric strengths of both the processes have a
similar thickness dependence as follows:
\begin{eqnarray}\label{De1}
\Delta\epsilon/\Delta\epsilon_{bulk}=1-\frac{\tilde a}{d},
\end{eqnarray}
where $\Delta\epsilon_{bulk}$ is the relaxation 
strength of the bulk sample, and $\tilde a$ is a constant independent of $d$.
The characteristic lengths fitted by Eq.(\ref{De1}) are 13.8$\pm$0.3 nm 
and 6.5$\pm$0.3 nm for the $\alpha$-process
in PVAc and the $\beta$-process in PMMA, respectively

\section{\bf Discussions}

We have successfully observed many properties of dynamics of the
$\alpha$-process in PVAc and the $\alpha$- and $\beta$-processes in
PMMA. Here, we will discuss some of the results 
observed in this study.

\subsection{\bf Thickness dependence of $T_{\alpha}$ in PVAc}

In the case of PVAc, it has been observed that 
the temperature $T_{\alpha}$ of the $\alpha$-peak
in the temperature domain at 100Hz decreases with decreasing film
thickness in a manner described by Eq.(\ref{Tg_Keddie}).
The thickness dependence of $T_{\alpha}$ has almost no 
$M_{\rm w}$ dependence in the $M_{\rm w}$ range from 1.2$\times$10$^5$ 
to 2.4$\times$10$^5$. 

In the case of the freely standing films of 
polystyrene, Mattsson {\it et al.} show that in a low $M_{\rm w}$ regime 
($M_{\rm w}\le $3.47$\times$10$^5$) the value of $T_{\rm g}$ observed 
by Brillouin light scattering has no molecular weight dependence and it
decreases gradually with decreasing film thickness in a similar way to
Eq.(\ref{Tg_Keddie})~\cite{Mattsson}. 
On the other hand, in a high $M_{\rm w}$ regime
($M_{\rm w}\ge $5.14$\times$10$^5$) the thickness dependence of 
$T_{\rm g}$ shows a distinct molecular weight dependence given by
Eq.(\ref{Tg_Forrest}) and it depends on the radius of gyration of the
polymer chain~\cite{Mattsson}. 

Although the existence of the interfacial interaction may lead to
the difference in the $d$ and $M_{\rm w}$ dependence of 
$T_{\rm g}$ between the freely standing films and the thin films supported
on substrate, our recent study shows that the temperature
$T_{\alpha}$ observed at 100Hz and 1kHz by dielectric relaxation
spectroscopy has a similar thickness and $M_{\rm w}$ dependence to
Eq.(\ref{Tg_Forrest}) for thin PS films 
supported on Al-deposited glass substrate~\cite{Fukao1,Fukao2}.  
Therefore, it is expected that the thickness and $M_{\rm w}$ dependence
of $T_{\alpha}$ in thin films supported on substrate can be compared
with that of $T_{\rm g}$ in freely standing thin films.
For this reason, the observed results on $T_{\alpha}$ in thin films of PVAc 
may correspond to the low $M_{\rm w}$ regime observed by Mattsson {\it et
al.} in the freely standing films of PS. In the regime, {\it confinement 
effect of polymer chains} within thin films is not dominant for the
reduction of $T_{\rm g}$ or $T_{\alpha}$, but {\it the finite size effects}
due to a length scale intrinsic to the glass transition may be dominant.
If we measure $T_{\alpha}$ for thin films of PVAc with still higher molecular
weights, however, $M_{\rm w}$ dependence of $T_{\alpha}$ $vs.$ $d$ may be 
observed in a similar way to $T_{\rm g}$ in the freely standing films of
PS.

\subsection{\bf Thickness dependence of $T_{\alpha}$ and $T_{\beta}$ in PMMA}

In PMMA, it can be expected that the
characteristic length scale of the $\beta$-process, if any, should be
smaller than that of the $\alpha$-process, 
because the characteristic time of the $\beta$-process at a fixed
temperature in the temperature range investigated here is much shorter 
than that of the $\alpha$-process. In this case,   
the thickness dependence of $T_{\alpha}$ should be
stronger than that of $T_{\beta}$ 
at a fixed frequency. However, the observed results do not obey this
expectation; as shown in Fig.11, for $d>d_{\alpha,c}$
($\approx d_{\beta,c}$), $T_{\alpha}$ remains
{\it almost independent of thickness} and $T_{\beta}$ {\it decreases slightly 
with decreasing film thickness}.

According to a layer model, as discussed in the 
literature~\cite{Keddie1,Fukao2,DeMaggio}, 
$T_{\rm g}$ and the dynamics of thin films are controlled by the
competition between the two different effects. Firstly, the existence
of a mobile layer near surfaces leads to the speed-up of the
dynamics of the $\alpha$-process (and also the $\beta$-process).
Within the mobile layer molecular motions are enhanced compared with
those in the bulk state. The origin of the enhancement of molecular
motion is supposed to be due to the higher density of chain ends of
polymer chains~\cite{Kajiyama} or to the restriction on cooperative motions
characteristic of the glass transition 
within the surface layer~\cite{Mattsson,Jerome}. 
In the latter case, the layer 
thickness of the mobile layer is of an order of the size of 
the cooperatively rearranging region (CRR)~\cite{Mattsson}. 
Secondly, the attractive interaction between
the polymers and the substrate leads to slowing-down of the dynamics of
the motion~\cite{Blum1,Blum2}. For example, the adsorption of polymer 
chains onto the
substrate causes $T_{\rm g}$ of the thin films to increase compared with 
that in the bulk state. 

In the case of thin films of PMMA, it has been reported that the
glass transition temperature strongly depends on the interaction 
between the polymers and the substrate. Keddie {\it et al.} show that
{\it the
$T_{\rm g}$ increases with decreasing film thickness} for SiO$_2$
substrate, while {\it it decreases with decreasing film thickness} for Au 
substrate~\cite{Keddie2}. 
Local thermal analysis measurements showed that $T_{\rm g}$ 
of thin films of PMMA supported on SiO$_x$ is by 7K higher for $d$=20nm 
than that for the bulk films, while $T_{\rm g}$ on SiO$_x$ with 
hexamethyldisilizane is by 10K lower for $d$=20nm~\cite{Pablo}.
In thin films of PMMA supported on Al deposited silicon
substrate, it is reported that $T_{\rm g}$ of thin films with $d$=40nm
is higher by 14$\pm$4 K than that of bulk PMMA~\cite{Grohens}.
These results suggest that the effect of the interaction between the
polymer and the substrate is sensitive to the condition of the surface
of the substrate in thin films of PMMA. 
In the $\alpha$-process in PMMA on Al-deposited substrate the mobile 
layer associated with cooperative length scale of the glass transition
tends to reduce $T_{\rm g}$ and $T_{\alpha}$, while the attractive 
interaction between the polymer and the substrate tends to increase
them.  As a result, the two effects cancel out and the observed value 
of $T_{\alpha}$ may {\it remain almost constant} for $d>d_{\alpha,c}$, as
shown in Fig.11(a). 
For the $\beta$-process, it is expected that the two effects no longer 
cancel out and the mobile layer has dominant effect on $T_{\beta}$.
In order to check the validity of the above idea, we should measure 
$T_{\alpha}$ and $T_{\beta}$ of PMMA thin films for a substrate which
has almost no attractive interaction between the polymer and the
substrate. In such a case, it is expected that as the thickness decreases
$T_{\alpha}$ decreases faster than $T_{\beta}$ does.

In the present measurements, it is observed that  for $d<d_{\alpha,c}$ a
drastic decrease in $T_{\alpha}$ begins with decreasing film thickness. 
To our surprise, $T_{\beta}$ also begins to decrease drastically with 
decreasing film thickness at almost the same film thickness
$d_{\beta,c}(\approx d_{\alpha,c})$. This indicates that there is a
strong correlation between the $\alpha$-process and the $\beta$-process
in PMMA thin films, although both the processes are located away from
each other in the dispersion map (Fig.10). Both the values of 
$d_{\alpha,c}$ and $d_{\beta,c}$ agree well with the radius of gyration
of the polymer chain ($R_{\rm g}\approx$18.6nm), and hence this behavior 
may be controlled by the polymeric nature of PMMA.

The drastic change at $d_{\alpha,c}$ implies the
crossover from the normal $\alpha$-process to other type of
dynamics. Recently, de Gennes proposed sliding motion, which can be 
activated via a soft skin layer near the surface instead of normal
$\alpha$-dynamics~\cite{deGennes1,deGennes2}. Below $d_c$, the sliding
motion may be activated in this case. The measurements of $M_{\rm w}$ 
dependence of $T_{\alpha}$ and $T_{\beta}$ for thin films of PMMA are
highly desirable in order to clarify the nature of the above behavior on 
$T_{\alpha}$ and $T_{\beta}$.

\subsection{Distribution of the relaxation times}

The distribution of the relaxation times of the $\alpha$-process
$\tau_{\alpha}$ in thin films of PVAc has been found to become broader
with decreasing film thickness. The width $w$ of the distribution in
PVAc with $d$=9.5nm is about 2.5 times as broad as that in bulk PVAc.
On the other hand, the width $w$ for the $\beta$-process in thin films
of PMMA depends on film thickness much more weakly, as shown in
Fig.8(a). The corresponding width $\Delta T_i$ in temperature domain has 
a similar thickness dependence, $i.e.$, the thickness dependence of $\Delta
T_{\beta}$ is much weaker than that of $\Delta T_{\alpha}$ in PMMA. From
these observations, it is found that the distribution of relaxation
times of the $\alpha$-process depends on film thickness much stronger
than that of the $\beta$-process. The intrinsic length scale of the
$\beta$-process is expected to be much smaller than that of the
$\alpha$-process, as deduced from Fig.10. The present observation on the 
thickness dependence of the 
distribution of the relaxation times is consistent with this expectation.

\subsection{Thickness dependence of $\Delta\epsilon$}

As shown in Fig.2, the dielectric loss peak due to the
$\alpha$-process becomes more depressed with decreasing film
thickness. 
This is consistent with the result that the relaxation strength
$\Delta\epsilon$ decreases with decreasing film thickness, as shown in
Fig.5. The thickness dependence of $\Delta\epsilon$ is 
found to obey Eq.(\ref{De1}). Here, we will try to give a possible
explanation on the decrease in $\Delta\epsilon$ with decreasing thickness.

If there is no interaction between dipole moments,
the relaxation strength $\Delta\epsilon_0$ is described as follows:
\begin{eqnarray}\label{Delta_e}
\Delta\epsilon_0\approx \frac{N\mu_0^2}{3k_{\rm B}T},
\end{eqnarray}
where $N$ is the dipole number density, $\mu_0$ is the dipole moment
of the relaxing unit, $k_{\rm B}$ is the Boltzmann constant, and $T$ is
the temperature. Now it is assumed that {\it $n$ neighboring dipole moments
are relaxed in a cooperative manner}. Then $N$ and $\mu_0$ should be
replaced by $N/n$ and $n\mu_0$ in Eq.(\ref{Delta_e}), respectively. As a 
result, we obtain the relaxation strength $\Delta\epsilon_n$ for 
such a cooperative motion in the following way:
\begin{eqnarray}
\Delta\epsilon_n\approx n\frac{N\mu_0^2}{3k_{\rm B}T}.
\end{eqnarray}
Here, it can be expected that the number $n$ is proportional to the
intrinsic length of the motion in question. Because cooperative motions
are restricted near surfaces or interfaces, the intrinsic length of the
motions may be reduced and $n$ is decreased. Therefore, $\Delta\epsilon$
of the region near surfaces and interfaces can be expected to be smaller 
than that of the bulk system. By averaging along the direction normal to
the film surface, the relaxation strength $\Delta\epsilon$ 
is expressed by
\begin{eqnarray}
\Delta\epsilon=\Delta\epsilon_{bulk}\left[1-\xi\left(1-
\frac{\Delta\epsilon_{surf}}{\Delta\epsilon_{bulk}}\right)\frac{1}{d}\right],
\end{eqnarray}
where $\xi$ is the effective thickness of the surface region,
$d$ is the overall thickness, and $\Delta\epsilon_{bulk}$ and 
$\Delta\epsilon_{surf}$ are the relaxation strength of the bulk 
region and the surface regions, respectively.
Here, the relation $\Delta\epsilon_{bulk}> \Delta\epsilon_{surf}$ is
satisfied.
On the basis of this simple picture, the observed thickness dependence 
of $\Delta\epsilon$ in Eq.(\ref{De1}) can be reproduced. The value of
$\tilde a$ is given by $\tilde 
a=\xi(1-\Delta\epsilon_{surf}/\Delta\epsilon_{bulk})$. 
Therefore, the observed result on $\Delta\epsilon$ in the present
measurements is consistent with the idea of cooperative motions.
 

\vspace*{-0.6cm}
\section{\bf Summary}
We made dielectric relaxation measurements for thin films of poly(vinyl
acetate) and poly(methyl methacrylate)
supported on Al deposited glass substrate. The results obtained in the
present measurements are summarized as follows:
\begin{enumerate}
\item
The temperature $T_{\alpha}$ corresponding to the peak in the dielectric 
loss due to the $\alpha$-process at 100Hz is measured as a function of
film thickness for thin films of PVAc with three different molecular
weights $M_{\rm w}$=1.2$\times$10$^5$, 1.8$\times$10$^5$ and
2.4$\times$10$^5$. It is found that $T_{\alpha}$ decreases gradually
with decreasing film thickness and the decreasing rate of $T_{\alpha}$
increases more and more in thinner films. Furthermore, $T_{\alpha}$ 
has almost no molecular weight dependence.
\item
The width of the distribution of the relaxation times of the
$\alpha$-process increases with decreasing film thickness, while
the relaxation strength of the $\alpha$-process decreases with
decreasing film thickness.
\item
The thickness dependence of  the temperatures $T_{\alpha}$ and
$T_{\beta}$ corresponding to the peaks in the dielectric loss due
to the $\alpha$- and $\beta$-processes at 40Hz, respectively, are
measured for thin films of PMMA with $M_{\rm w}$=4.9$\times$10$^5$. The 
temperature $T_{\alpha}$ remains almost constant as the thickness is
decreased down to a critical thickness $d_{\alpha,c}$, at which
point it begins to decrease with thickness. The value of
$d_{\alpha, c}$ is about 20nm. On the other hand, the temperature
$T_{\beta}$ decreases slightly with decreasing film thickness from the
bulk to a critical thickness $d_{\beta,c}$. Below $d_{\beta,c}$
$T_{\beta}$ decreases more rapidly with decreasing film thickness.
The value of $d_{\beta,c}$ is almost equal to $d_{\alpha,c}$.
\item
The widths $\Delta T_{\alpha}$ and $\Delta T_{\beta}$ of the peaks in
dielectric loss in the temperature domain are measured for both the
$\alpha$- and $\beta$-processes in thin films of PMMA with 
$M_{\rm w}$=4.9$\times$10$^5$. As the thickness is decreased, the width
$\Delta T_{\alpha}$ increases much faster than $\Delta T_{\beta}$. This
indicates that the broadening of the distribution of relaxation times of
the $\alpha$-process is much more sensitive to the change in thickness
than that of the $\beta$-process.
\end{enumerate}

In this paper, several length scales characteristic of the thickness
change in $T_{\alpha}$, $\Delta T_{\alpha}$, $T_{\beta}$, $\Delta
T_{\beta}$, $\Delta\epsilon$, and $w$ have been introduced.
The physical quantities such as $\Delta T_{\alpha}$, $\Delta T_{\beta}$, 
$w$ and 
$\Delta\epsilon$ have been found to be described by the linear equation
of the reciprocal of $d$. This result suggests that a layer model be
applicable to explain the thickness dependence, as shown in IV.D. In such
cases, the characteristic lengths $a$ and  $\tilde a$ are correlated 
with the thickness of the surface layer $\xi$, which can be expected to
be comparable to the size of the CRR. On the other hand, the crossover
thicknesses $d_{\alpha,c}$ and $d_{\beta,c}$ are almost equal to the
radius of gyration of the polymer chain and are expected to be molecular 
weight dependent. These crossover thicknesses may be strongly associated 
with dynamics of polymer chains. 
In order to understand the present experimental results, a model will
be required, where the nature of polymer dynamics is taken
into account as well as the intrinsic nature of the glass transition.  

\section*{Acknowledgments}
The work was partly supported by a Grant-in-Aid from the Ministry 
of Education, Science, Sports and Culture of Japan.

\begin{minipage}{8.5cm}
\begin{table}
\caption{The weight averaged molecular weight $M_{\rm w}$, the ratio of 
$M_{\rm w}$ to the number averaged molecular weight $M_{\rm n}$, and radius of gyration of the polymers $R_{\rm g}$ 
used in this studies}
\begin{tabular}{lccc}
 & $M_{\rm w}$ & $M_{\rm w}/M_{\rm n}$ & $R_{\rm g}$(nm)$^{\dagger}$ \\\hline
PVAc & 124,800 & 2.37 & 10.1 \\
 & 182,000 & 2.95 & 12.2 \\
 & 237,100 & 2.64 & 12.8 \\
PMMA & 490,200 & 4.11 & 18.6
\end{tabular}
$^{\dagger}$ The values of $R_{\rm g}$ are obtained from
Ref.\cite{Physical}.
\end{table}
\end{minipage}

\vspace*{0.8cm}
\begin{figure}\label{fig:fig1}
\epsfxsize=8.5cm 
\centerline{\epsfbox{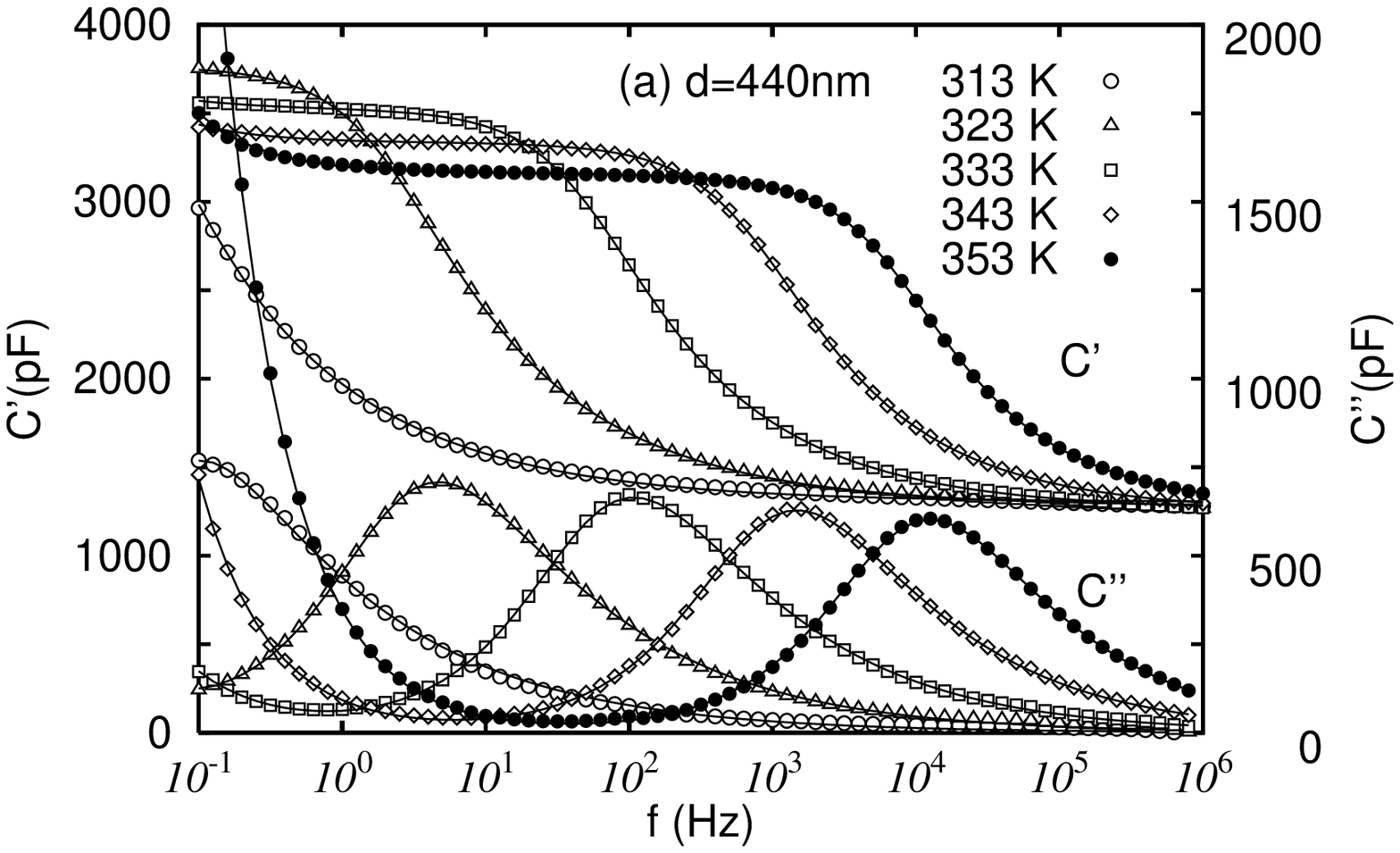}}
\epsfxsize=8.5cm 
\centerline{\epsfbox{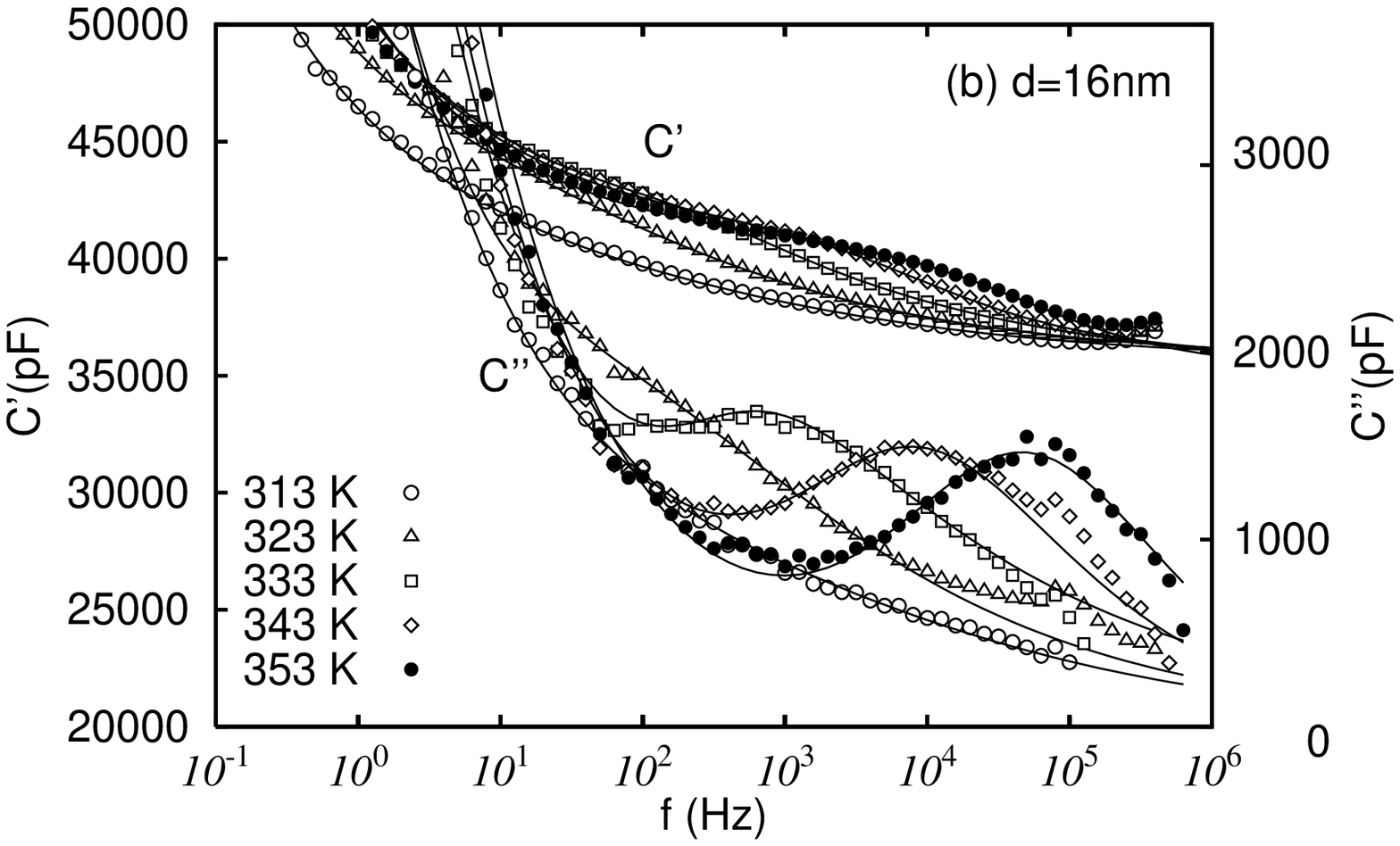}}
\begin{minipage}{8.5cm}
\caption{
The dependence of the complex electric capacitance on the logarithm of
frequency at various temperatures for PVAc with $M_{\rm w}=1.8\times 10^5$:
(a) d=440nm, (b) d=16nm. Solid curves are calculated by Eq.(3).
}
\end{minipage}
\end{figure}
\vspace{0.2cm}

\vspace*{-0.3cm}
\begin{figure}
\epsfxsize=8.5cm 
\centerline{\epsfbox{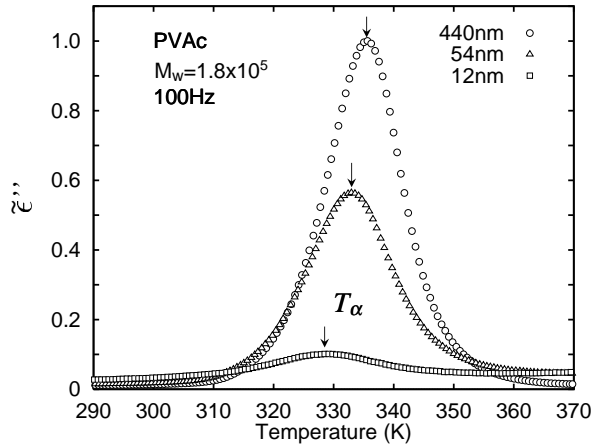}}
\vspace{0.0cm}
\begin{minipage}{8.5cm}
\caption{
The temperature dependence of the imaginary part of the complex
 dielectric constant at 100Hz for PVAc with $M_{\rm w}$=1.8$\times$10$^5$ 
and three different thicknesses: $\circ$ corresponds to 440nm,
 $\triangle$ to 54nm, and $\Box$ to 12nm. The value of the vertical axis
 is normalized with the maximum value for the bulk sample. The arrows
 indicate the values of $T_{\alpha}$ at 100Hz.
}
\end{minipage}
\label{fig:fig2}
\end{figure}
\vspace{0.2cm}

\vspace*{-0.3cm}
\begin{figure}
\epsfxsize=8.5cm 
\centerline{\epsfbox{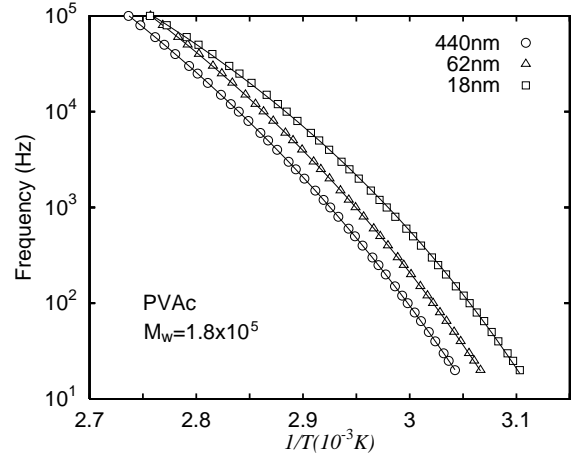}}
\vspace{0.0cm}
\begin{minipage}{8.5cm}
\caption{
The relationship between the temperature $T_{\alpha}$ and the 
frequency $f$ at which the
dielectric loss possesses a maximum due to the $\alpha$-process for 
PVAc with $M_{\rm w}$=1.8$\times$10$^5$ and three different thicknesses:
$\circ$ corresponds to 440nm, $\triangle$ to 62nm, and $\Box$ to 18nm.
Solid curves are obtained by fitting the data to the VFT equation.
}
\end{minipage}
\label{fig:fig3}
\end{figure}
\vspace{0.2cm}

\vspace*{0.0cm}
\begin{figure}
\epsfxsize=8.5cm 
\centerline{\epsfbox{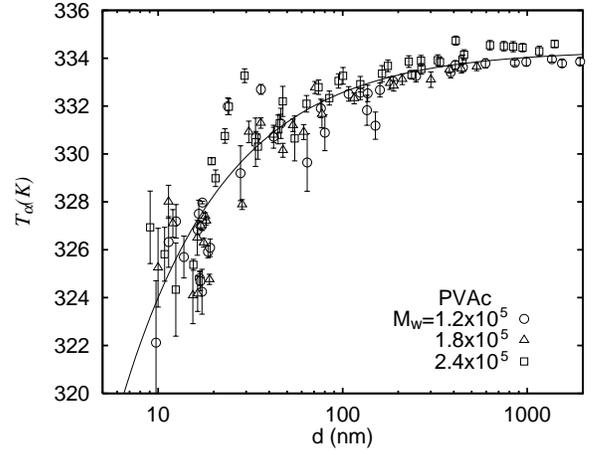}}
\vspace{0.4cm}
\begin{minipage}{8.5cm}
\caption{
Thickness dependence of the temperature $T_{\alpha}$ for PVAc with
three different molecular weights: $\circ$ corresponds to 
$M_{\rm w}$=1.2$\times$10$^5$, $\triangle$ to 1.8$\times$10$^5$, and
$\Box$ to 2.4$\times$10$^5$. The temperature $T_{\alpha}$ is a
temperature at which the imaginary component of the complex dielectric
constant possesses a maximum value at 100Hz.}
\end{minipage}
\label{fig:fig4}
\end{figure}
\vspace{0.2cm}

\vspace*{0.0cm}
\begin{figure}
\epsfxsize=8.5cm 
\centerline{\epsfbox{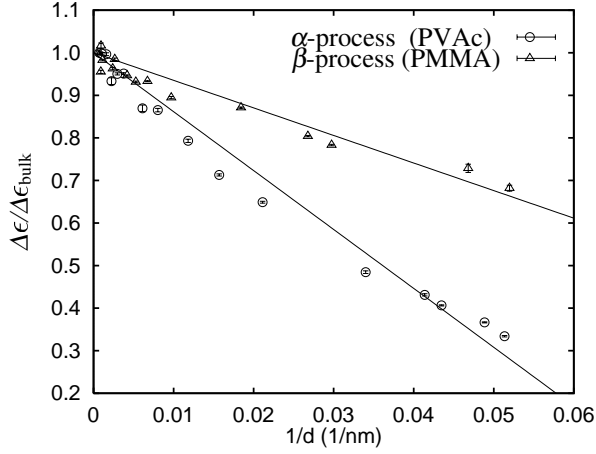}}
\vspace{0.0cm}
\begin{minipage}{8.5cm}
\caption{Thickness dependence of the relaxation strength normalized
with the value of the bulk sample. Open circles correspond to the result 
obtained at 333 K for the $\alpha$-process in PVAc with 
$M_{\rm w}$=2.4$\times$10$^5$ and open
triangles to the result obtained at temperatures at which the dielectric 
loss peak is located at 300Hz for the $\beta$-process in PMMA 
with $M_{\rm w}$=4.9$\times$10$^5$.
Solid lines are obtained by Eq.(\protect\ref{De1}). The characteristic
lengths are 13.84$\pm$0.34 nm and 6.48$\pm$0.31 nm for the $\alpha$-process
in PVAc and the $\beta$-process in PMMA, respectively
}
\end{minipage}
\label{fig:fig5}
\end{figure}
\vspace{0.2cm}

\vspace*{0.2cm}
\begin{figure}
\epsfxsize=8.5cm 
\centerline{\epsfbox{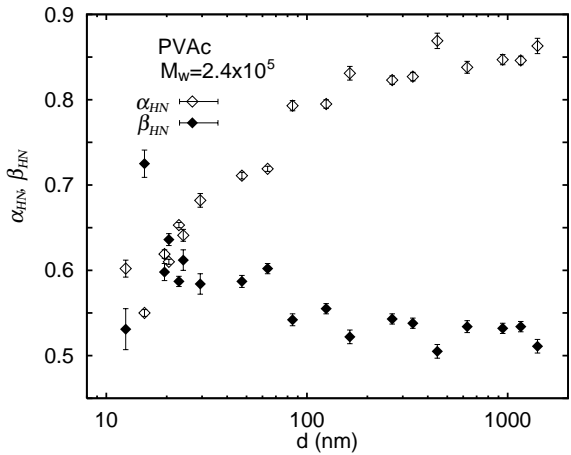}}
\vspace{0cm}
\begin{minipage}{8.5cm}
\caption{The thickness dependecne of the shape parameters 
$\alpha_{\rm HN}$ and
$\beta_{\rm HN}$ of the HN equation obtained by fitting the 
observed data at
333K for PVAc with $M_{\rm w}$=2.4$\times$10$^5$:
open diamonds correspond to $\alpha_{\rm HN}$ 
and full diamonds to $\beta_{\rm HN}$.}
\end{minipage}
\label{fig:fig6}
\end{figure}
\vspace{0.2cm}

\begin{figure}
\epsfxsize=8.5cm 
\centerline{\epsfbox{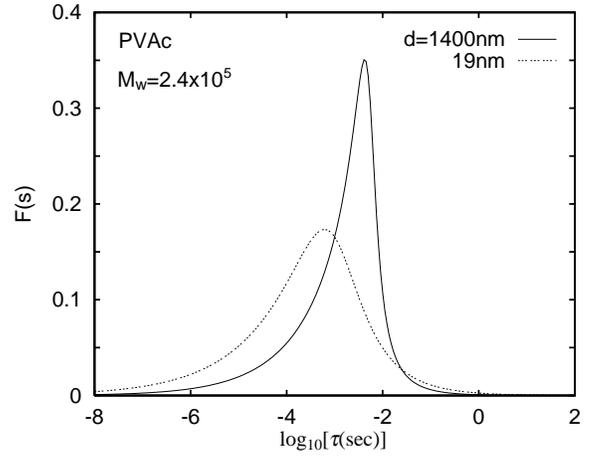}}
\vspace{0cm}
\begin{minipage}{8.5cm}
\caption{The distribution function $F(s)$ of the relaxation times of the 
$\alpha$-process at 333K for PVAc with $M_{\rm w}$=2.4$\times$10$^5$.
The values are calculated by Eq.(5) with the best-fit parameters of the 
HN equation. The solid curve stands for the result of $d$=1400nm and the 
 dotted curve for that of $d$=19nm.}
\end{minipage}
\label{fig:fig7}
\end{figure}
\vspace{0.2cm}


\vspace*{0.0cm}
\begin{figure}
\epsfxsize=8.5cm 
\centerline{\epsfbox{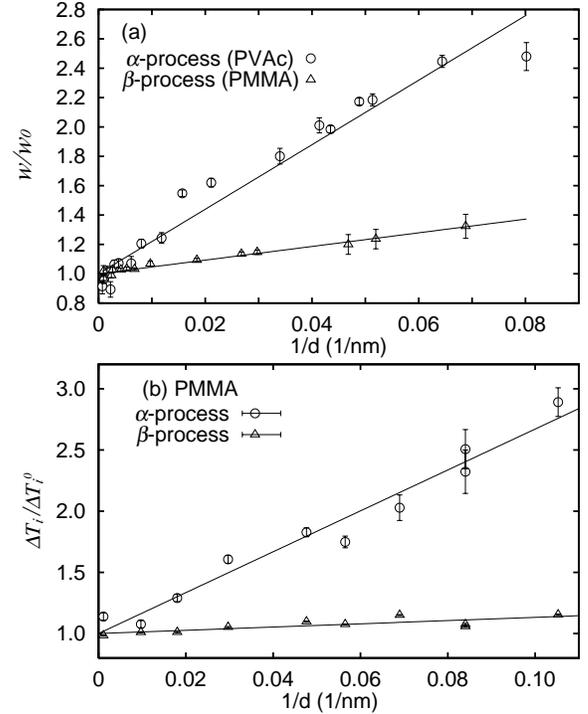}}
\vspace{-3.5cm}
\begin{minipage}{8.5cm}
\caption{(a) The thickness dependence of the normalized full width at
 the half maximum $w/w_0$ for the $\alpha$-process in PVAc 
with $M_{\rm w}$=2.4$\times$10$^5$ ($\circ$), and the $\beta$ in PMMA 
with $M_{\rm w}$=4.9$\times$10$^5$ ($\triangle$). The width for PVAc is
 evaluated at 333K and that for PMMA is at the temperature at which the
 $\beta$-peak in the dielectric loss is located at 300Hz. (b) The
 thickness dependence of the width $\Delta T_{\alpha}$ and $\Delta T_{\beta}$ 
 of the $\alpha$-, $\beta$-peaks in
 the dielectric loss at 40Hz in the temperature domain. The width is
 normalized with that of the bulk sample.}
\end{minipage}
\label{fig:fig8}
\end{figure}
\vspace{0.2cm}

\vspace*{-0.5cm}
\begin{figure}
\epsfxsize=8.8cm 
\centerline{\epsfbox{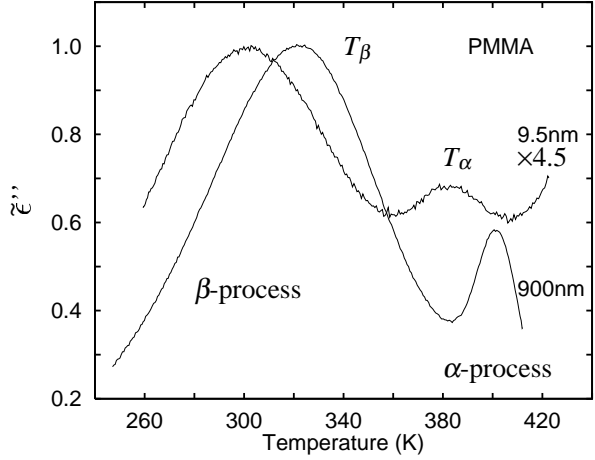}}
\vspace{0.3cm}
\begin{minipage}{8.5cm}
\caption{
The temperature dependence of the imaginary part of the complex
dielectric constant at 40Hz for PMMA with the thickness of 9.5nm and
900nm. The vertical axis is normalized with the maximum 
value for the bulk sample. The value for $d$=9.5nm is magnified
by a factor of 4.5.  
}
\end{minipage}
\label{fig:fig9}
\end{figure}
\vspace{0.2cm}

\vspace*{-0.2cm}
\begin{figure}
\epsfxsize=8.8cm 
\centerline{\epsfbox{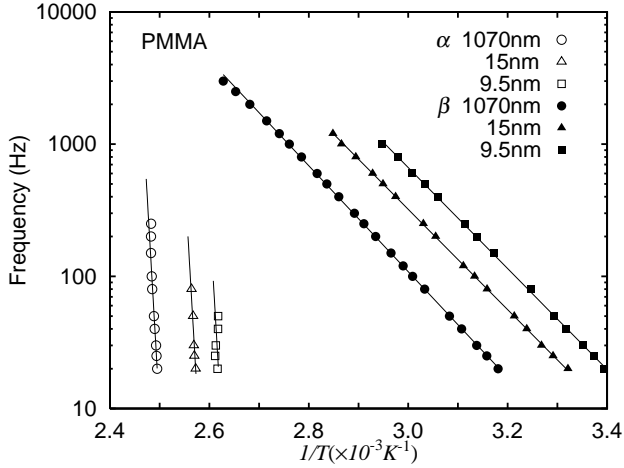}}
\vspace{0.3cm}
\begin{minipage}{8.5cm}
\caption{Dispersion map of PMMA for various thicknesses. Circles
 correspond to $d$=1070nm, triangles to $d$=15nm, and boxes to
 $d$=9.5nm. Full and open symbols are for the $\alpha$-process and the
 $\beta$-process, respectively.}
\end{minipage}
\label{fig:fig10}
\end{figure}
\vspace{0.2cm}

\vspace*{-0.6cm}
\begin{figure}
\epsfxsize=8.8cm 
\centerline{\epsfbox{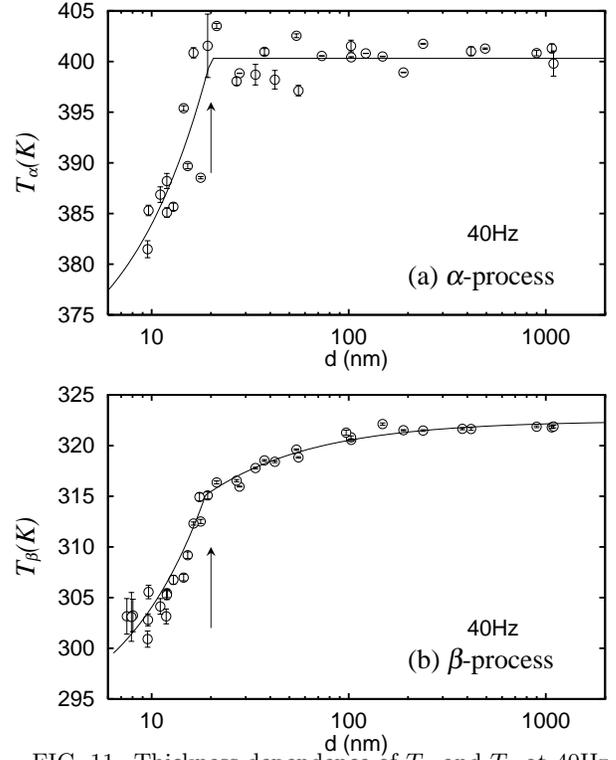}}
\vspace{-0.6cm}
\begin{minipage}{8.5cm}
\caption{Thickness dependence of $T_{\alpha}$ and $T_{\beta}$ at 40Hz
 for PMMA. The arrows indicate the critical thickness $d_c$.}
\end{minipage}
\label{fig:fig11}
\end{figure}
\vspace{0.2cm}

\vspace*{-0.9cm}
\begin{figure}\label{fig:fig12}
\epsfxsize=8.5cm 
\centerline{\epsfbox{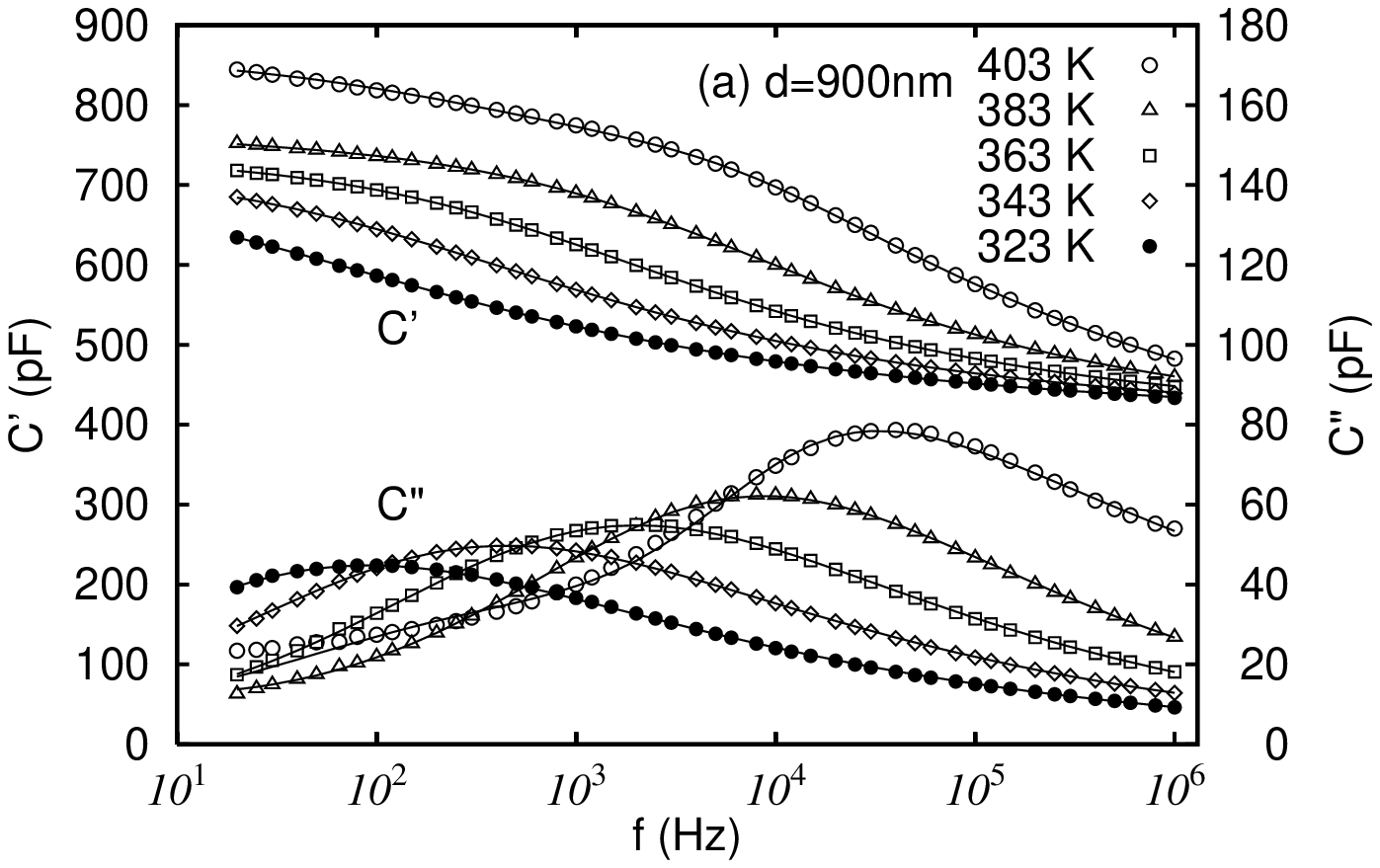}}
\epsfxsize=8.5cm 
\centerline{\epsfbox{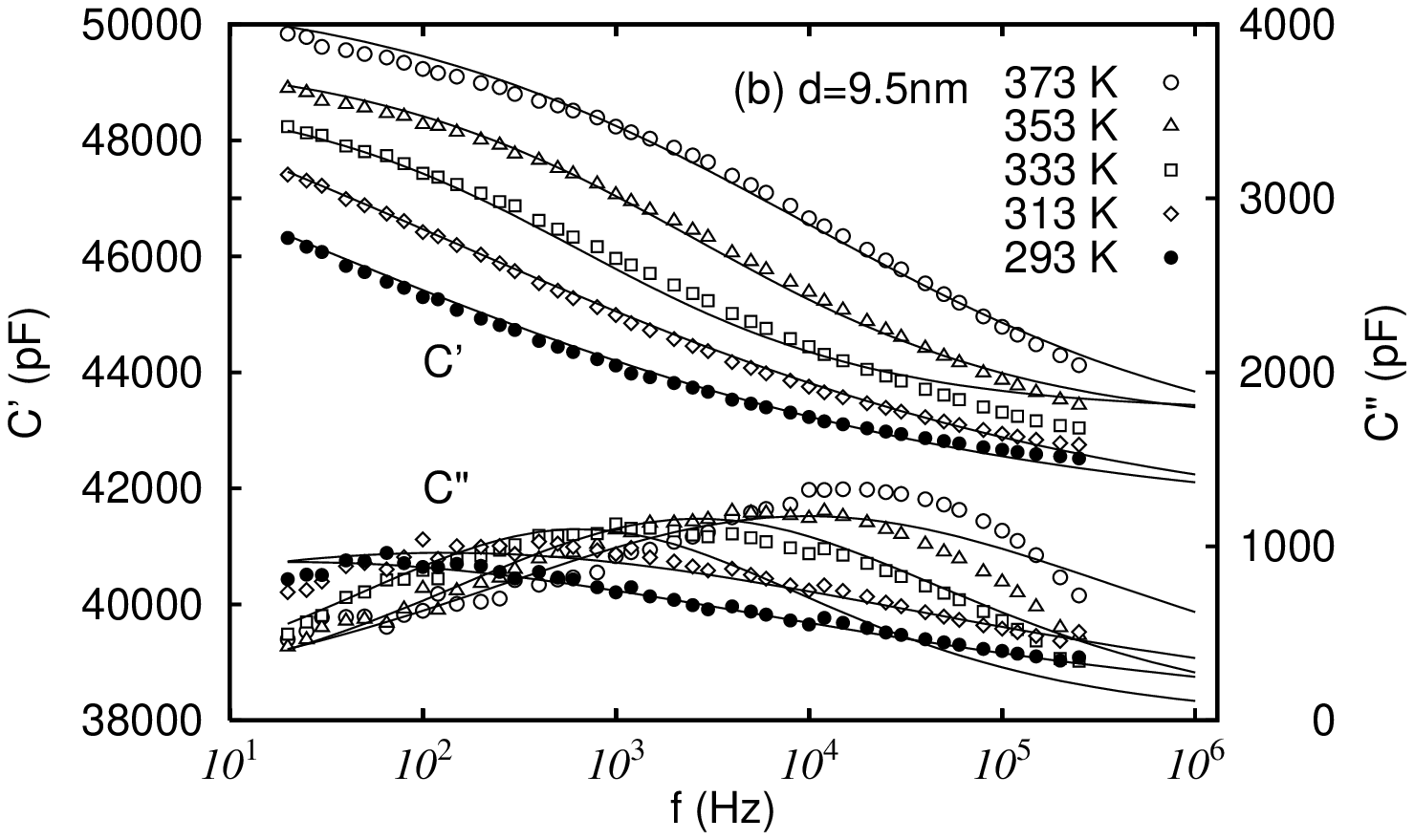}}
\begin{minipage}{8.5cm}
\caption{
The dependence of the complex electric capacitance on the logarithm of
frequency at various temperatures for PMMA with $M_{\rm w}$=4.9$\times 10^5$:
(a) d=900nm, (b) d=9.5nm. Solid curves are calculated by the sum of two
 HN equations.
}
\end{minipage}
\end{figure}
\vspace{0.2cm}
\newpage
\end{multicols}

\end{document}